\documentclass[numberedappendix,twocolumn,onecolappendix]{openjournal}

\usepackage{newtxtext,newtxmath}

\usepackage[T1]{fontenc}

\DeclareRobustCommand{\VAN}[3]{#2}
\let\VANthebibliography\thebibliography
\def\thebibliography{\DeclareRobustCommand{\VAN}[3]{##3}\VANthebibliography}

\usepackage{graphicx}	
\usepackage{amsmath}	

\newcommand{\Msun}{M_\odot}
\newcommand{\orcidauthor}[3]{\author{\href{http://orcid.org/#1}{#2$^{#3}$}}}
\newcommand{\cutdefinition}{All cuts involve a halo mass selection ($0.5\times10^{12}\Msun<M_h<2.0\times10^{12}\Msun$).  The Strict and Extended cuts also involve selecting for M31-like neighbors to approximate the local environment. For detailed cut definitions, see Section \ref{s:mw_analog}.}

\usepackage{hyperref}
\usepackage{amsmath}
\usepackage[x11names]{xcolor}

\hypersetup{
    colorlinks=true,
    linkcolor=blue,
    citecolor=blue,
    filecolor=magenta,      
    urlcolor=blue,
    pdfpagemode=FullScreen,
    }

\begin{document}

\title[Origin of High-Velocity Particles in the MW]{Where do High-Velocity Dark Matter Particles Come From in the Milky Way?}

\author{Aidan DeBrae$^{1}$}

\orcidauthor{0000-0002-2517-6446}{Peter Behroozi}{1}

\orcidauthor{0000-0001-7107-1744}{Nicolás Garavito-Camargo}{2, 1}

\affiliation{$^{1}$Department of Astronomy and Steward Observatory, University of Arizona, 933 N Cherry Ave, Tucson, AZ, 85721, USA}

\affiliation{$^{2}$Center for Computational Astrophysics, Flatiron Institute, 162 5th Ave, New York, NY 10010, USA}

\date{Accepted XXX. Received YYY; in original form ZZZ}



\begin{abstract}
High-velocity particles ($v>v_\mathrm{esc}$) in the Milky Way are rare but nonetheless important to characterize due to their impact on dark matter (DM) direct detection experiments.  We select halos similar in mass to the Milky Way in a large-volume dark matter simulation and measure the incidence of high-velocity particles, finding that an average fraction $\sim 1.3\times 10^{-5}$ of the DM particles have $\Delta v > 600$ km/s within 5--11 kpc of the halo centers.  However, some systems have dramatically higher fractions.  Milky Way-like systems with high-speed satellites can have high-velocity DM fractions of order $\sim 100\times$ higher than average.  The environment also affects high-velocity DM fractions; massive nearby halos ($>10^{13}\Msun$) can boost high-velocity DM density by $\sim 10\times$, although there is little effect for nearby Andromeda-like systems.  We confirm previous predictions from zoom-in simulations that the high-velocity particles in the Milky Way with heliocentric speeds $>700$ km/s primarily originate from the Large Magellanic Cloud, and provide a table of expected high-velocity DM densities at different heliocentric velocity thresholds.
\end{abstract}




\section{Introduction}

Galaxies, including our own Milky Way, reside in the centers of extended dark matter halos in the $\Lambda$CDM paradigm. However, the nature of dark matter is still uncertain at present, with several candidate hypotheses (e.g., weakly-interacting massive particles, axions, and primordial black holes) all lacking firm detections (see \citealt{Billard22} for a review). This has spurred significant investment in dark matter direct detection experiments, especially for weakly-interacting particles (see \citealt{Schumann19} for a review).

Detection depends on the product of local dark matter density (a function of the halo mass, concentration, and substructure) with the collision cross section \citep[see][and references therein]{Green17}.  For multiple types of dark matter (e.g., inelastic DM or light DM), high velocity particles have much larger collision cross sections; high-velocity particles are also important for directionally-sensitive experiments \citep{Kuhlen10}.  Yet, commonly-used dark matter particle velocity distributions fall off rapidly near the halo escape velocity (e.g., the Maxwell--Boltzmann distribution or the \citealt{Mao13} model). In the case of the Milky Way, the escape velocity at the solar position is estimated to be 500--550 km/s \citep{Koppelman21}.

In past work, we have shown that dark matter halos commonly contain transient particles that exceed the escape velocity \citep{Behroozi13}, which could in turn impact the interpretation of existing dark matter searches.  In this previous work, halo environment was identified as a strong predictor of the number of unbound particles---for example, close proximity to a massive neighboring halo results in a higher fraction of high velocity particles.  This is because DM particles orbit past the virial radii of halos, and the high velocity dispersion of a massive halo would mean that there would be a continuous flux of high-velocity particles through all other nearby halos.  Given that the Milky Way is in relatively close proximity to a larger halo (M31), we were motivated to consider whether the presence of M31 could affect the particle velocity distribution at the center of the Milky Way.

More recently, \cite{Besla19} found in non-cosmological simulations that the LMC could be a source of extremely high-velocity particles (700--900 km/s) due to the LMC's velocity with respect to the orbital velocity of the Solar System around the Milky Way's center (see also \citealt{Donaldson22}).  Hence, we were also motivated to investigate whether cosmological simulations of Milky Way halos with LMC-like substructure could also have significant excesses of high-velocity particles, and whether this effect was stronger than the environmental effect mentioned above.

Similar to this paper, \cite{Orlik23} investigated how the presence of the LMC might influence direct detection experiments. Their findings with zoom-in simulations showed (as we also find) that DM particles in the solar region are shifted to higher-speed distributions when considering the effects of the LMC.  The present study differs in that we use a large-scale cosmological simulation (\textit{VSMDPL}), which allows for a much larger sample size ($\sim$ 30$\times$) to better average over the effects of orbital variance.  As well, the cosmological simulations contain halos in a wide range of large-scale environments, allowing us to measure the relative impact of being near a massive neighboring halo.  

We describe the methodology in Section \ref{s:methods}, present results in Sections \ref{s:results} and \ref{s:LMC}, link these results to previous findings in Section \ref{s:discussion}, and conclude in Section \ref{s:conclusions}. When discussing halo masses, we adopt the virial overdensity criterion of \cite{mvir_conv}.  Where not otherwise specified, we classify ``high velocity'' dark matter particles as those with velocities exceeding 600 km/s relative to the halo center.

\section{Methodology}
\label{s:methods}

\subsection{Simulation}\label{sim}

We use halos from the VSMDPL (Very Small MultiDark Planck; \citealt{Klypin16,RP16}) N-body simulation.  
The box size of this simulation is $160$ Mpc/$h$, and it contains $3840^{3}$ particles, each of mass $6.2\times10^6 \: \Msun/h$. The initial redshift of this simulation is $z=150$. The simulation assumes a flat $\Lambda$CDM cosmology with parameters $ \Omega_M=0.307$, $n=0.96$, and $\sigma_8=0.82$; we adopt $h=0.68$, consistent with \citep{Planck15} constraints. Halos and merger trees for the simulation were computed using the \textsc{Rockstar} halo finder and \textsc{Consistent Trees} merger tree codes \citep{Behroozi13_Rockstar, Behroozi13_CMergers}. In \textit{VSMDPL}, there are nearly 30,000 central Milky Way (MW) mass halos ($M_h = 0.5-2.0 \times10^{12}\Msun$), each having about 3000 particles within 10 kpc of their centers, so this simulation is ideal for studying how the incidence of rare high-velocity particles changes with environment.  For 
halo velocities, we use the \textsc{Rockstar} default of averaging the dark matter particle velocities within 10\% of $R_\mathrm{vir}$.

\subsection{Selecting Milky Way-Like Halos}
\label{s:mw_analog}
We first must select halos from \textit{VSMDPL} that are analogous to the Milky Way (MW). Currently, the MW halo is estimated to have a mass of about $0.5-2.0 \times10^{12} \Msun$ \citep{Wang20}, so we select all central halos (i.e., those not within the virial radius of any larger halo) within this mass range. The nearest massive cluster to the MW is the Virgo Cluster at $16.5$~Mpc away \citep{Mei07}, so we additionally require that MW analog halos should have no neighboring halo greater than $1.0\times10^{14} \Msun/h$ within $10$~Mpc. 

\subsection{Selecting MW--M31 Analogs}

M31 is hosted by the nearest larger halo within the Local Group. Thus, we use the properties of M31 as constraints to find MW--M31 analogs in our simulation. Observations place the M31 halo mass at about $3\times10^{12} \: M_{\sun}/h$, with an uncertainty of about 0.5~dex \citep[e.g.,][]{Patel22}. Measurements of the radial and tangential velocity are $\sim -109$~km/s and $\sim 17$ km/s, respectively \citep{VDM12}. 

\begin{figure}
\includegraphics[width=\columnwidth]{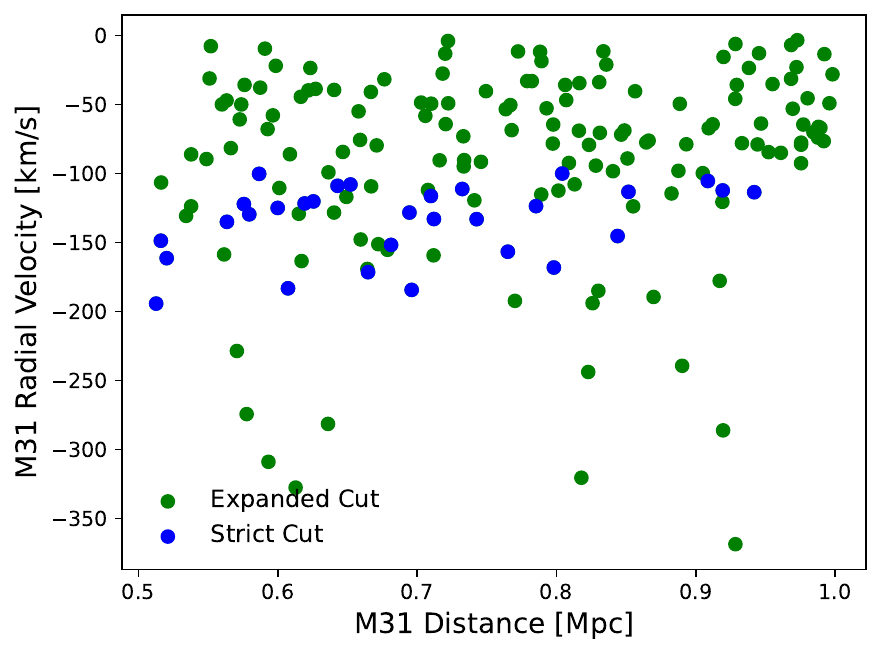}
    \caption{Radial velocities vs.\ distance for M31--like halos near MW-mass halos in the \textit{VSMDPL} dark matter simulation. M31--like halos from the Strict catalog have radial velocities which are selected to be $-(100-200)$ km/s, whereas those from the Expanded catalog are allowed to have a larger range of radial velocities for greater statistical power. The distribution of M31--like halo distances are similar for both catalogs.}
    \label{fig:M31_distVSradV}
\end{figure}

Because these observational constraints are so precise, it is difficult to obtain a sufficient statistical sample of exact matches from the simulation. Instead, we build sets of analogs with more relaxed cuts so that there are larger statistical samples for comparison. We hence select MW-like halos in three different ways, all (as above) requiring the MW-like halo to be a central halo with a mass of $0.5 < M_h/(10^{12}\Msun) < 2$ and no $>10^{14}\Msun/h$ cluster within a distance of 10 Mpc.  The additional criteria for each cut are:
\begin{itemize}
    \item \textbf{Strict Cut}: Larger M31--mass halo ($0.95 < M_h/(10^{12}\Msun) < 9.5$) at a  distance between $0.5-1.0$ Mpc,  with relative radial velocity $-200 < v_r/\textrm{(km/s)} < -100$ and tangential velocity $0 < v_t/\textrm{(km/s)} < 100$. There are 30 halos in the Strict catalog, with an average mass of $1.14\times10^{12} \Msun$. 
    \item \textbf{Expanded Cut}: Larger M31--mass halo at a distance between $0.5-1.0$~Mpc, with with relative radial velocity $-400 < v_r/\textrm{(km/s)} < 0$ and tangential velocity $0 < v_t/\textrm{(km/s)} < 500$. There are 175 halos in the Expanded catalog, with an average mass of $1.07\times10^{12} \Msun$. 
    \item \textbf{Full Cut}: No constraints on M31-like neighbors. There are 28305 halos in the Full catalog, with an average mass of $9.38\times10^{11} \Msun$. 
\end{itemize}
The result is three catalogs of MW analogs (Full, Expanded, and Strict), in which the properties of the neighboring halo get closer and closer to the observational constraints on the MW-M31 system for the Expanded and Strict catalogs. Figure \ref{fig:M31_distVSradV} shows the distribution of distances and radial velocities for the Expanded and Strict catalogs. 

 Because the Earth sits approximately $8$ kpc from the center of the halo, we select particles within $5-11$ kpc of halo centers for comparison.  This is similar to the particle distances selected in \cite{Orlik23}; halos in the Full catalog have on average 3300 particles (with a standard deviation of 1000) within this distance range. 

 \section{Results}\label{s:results}

\subsection{Impact of M31}

\begin{figure}
	\includegraphics[width=\columnwidth]{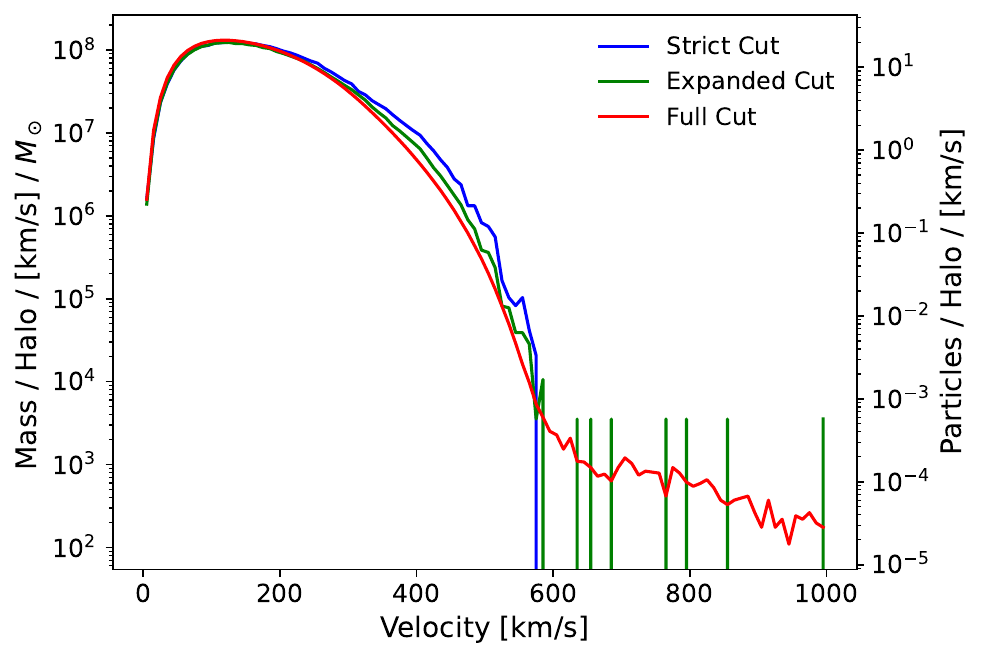}
    \caption{The presence of M31 does not substantially affect the particle velocity distribution near the Earth in the MW.  This figure shows average halocentric particle velocity distributions for Milky Way--like central halos in the \textit{VSMDPL} simulation, within 5--11 kpc of the halo center.  These are expressed in units of mass per halo per velocity bin width.  Small differences in the velocity distribution at 400--600 km/s arise because of $\sim 21\%$ differences in average masses across the different selection cuts (see text).  \cutdefinition{}}
    \label{fig:veldist}
\end{figure}

Fig.~\ref{fig:veldist} shows the halocentric velocity distributions for each set of DM halos defined in section \ref{s:mw_analog}. These results suggest there is not a substantial difference in the particle velocity distribution regardless of whether or not an M31--like halo is present.

\begin{figure}
	\includegraphics[width=\columnwidth]{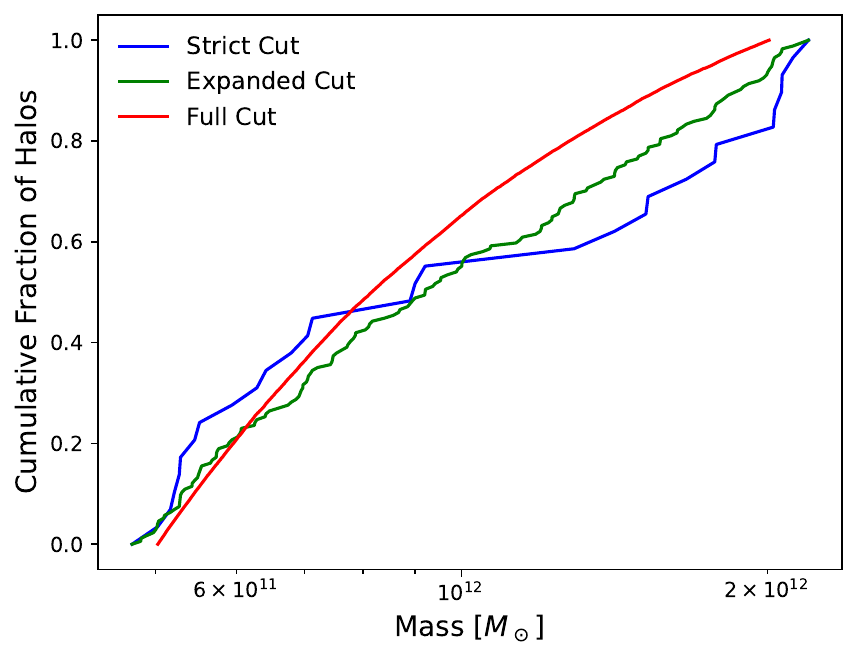}
    \caption{Cumulative distribution of halo masses in each catalog. Average masses are highest in the Strict catalog, followed by the Expanded, and lastly the Full catalog. The average mass is in the range of $0.9-1.2 \times10^{12} \Msun$ for all catalogs.}
    \label{fig:mwMassDistribution}
\end{figure} 

\begin{figure}
    \vspace{-7mm}
	\includegraphics[width=\columnwidth]{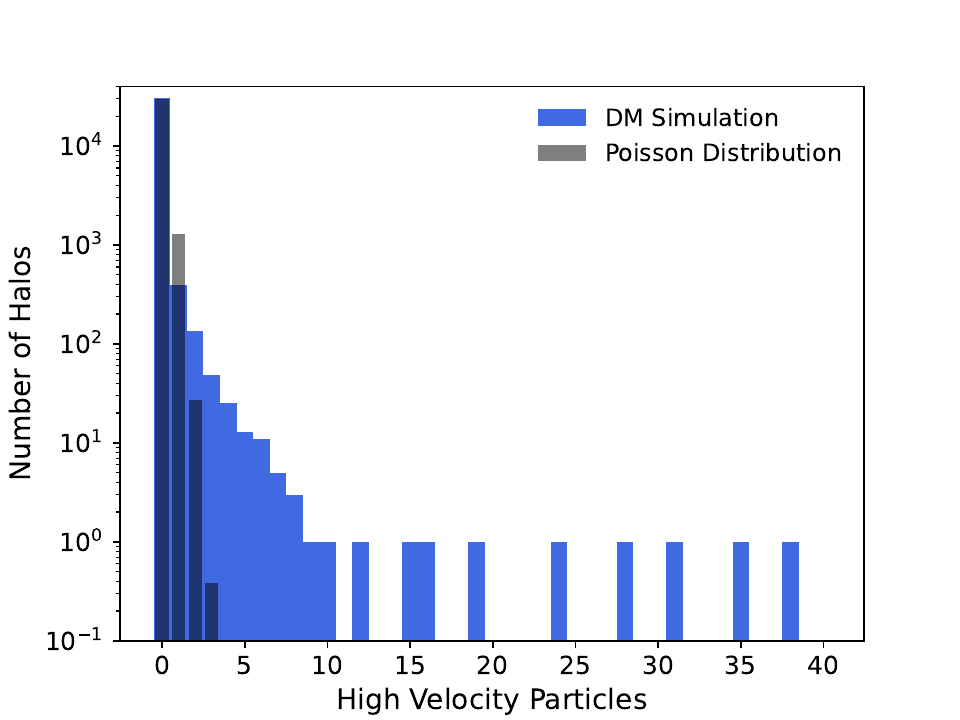}
    \caption{Histogram for counts of high-velocity particles ($v>600$ km/s) in the solar neighborhood in individual halos in the \textit{VSMDPL} simulation (\textit{blue bars}).  The dark shaded bars show the expected histogram if the number of high-velocity particles per halo obeyed a Poisson distribution.  Dark matter halos show an extended tail to large high-velocity particle counts, meaning that some halos have a much higher probability of hosting high-velocity particles than others.}
    \label{fig:poisson}
\end{figure}

Minor variations are seen for the incidence of 400$-$600 km/s particles, which are most common in the Strict catalog.  These are most likely due to small differences in the average masses of the different catalogs, which are $\sim$21\% larger for the Strict catalog than for the Full catalog (Fig.\ \ref{fig:mwMassDistribution}). The Expanded and Full catalogs of MW-analog halos have clear high-velocity tails, whereas halos in the Strict Cut do not contain any high velocity particles. According to Fig.~\ref{fig:veldist}, the incidence of high-velocity particles is similar for both the Expanded and Full catalogs; 4\% of sampled halos in those catalogs contain at least a single high-velocity particle. Because the Strict catalog contains so few halos, it is then not unusual that we do not observe any high velocity particles within this sample ($p>0.5$, assuming the distribution in Fig.\ \ref{fig:poisson}). Hence, our best estimate of high-velocity particle incidence in MW--M31 analog systems comes from the Expanded Cut, in which the presence of an M31 analog does not result in a strong difference in the number of high-velocity particles.

\subsection{Environmental and Substructure Origins of High-Velocity Particles}

Although rare, Fig.\ \ref{fig:veldist} makes it clear that a small fraction of central halos do contain high-velocity particles, including some with $v\gg v_\mathrm{esc}$. Fig.\ \ref{fig:poisson} shows a histogram of the number of high-velocity particles in the full catalog, showing that there is a non-Poissonian tail to very large numbers of high-velocity particles.  Hence, it becomes interesting to understand why certain halos have very large numbers of high-velocity particles, to determine whether the Milky Way would be likely to have a significant excess of high-velocity particles or not. 

\begin{figure}
	\includegraphics[width=\columnwidth]{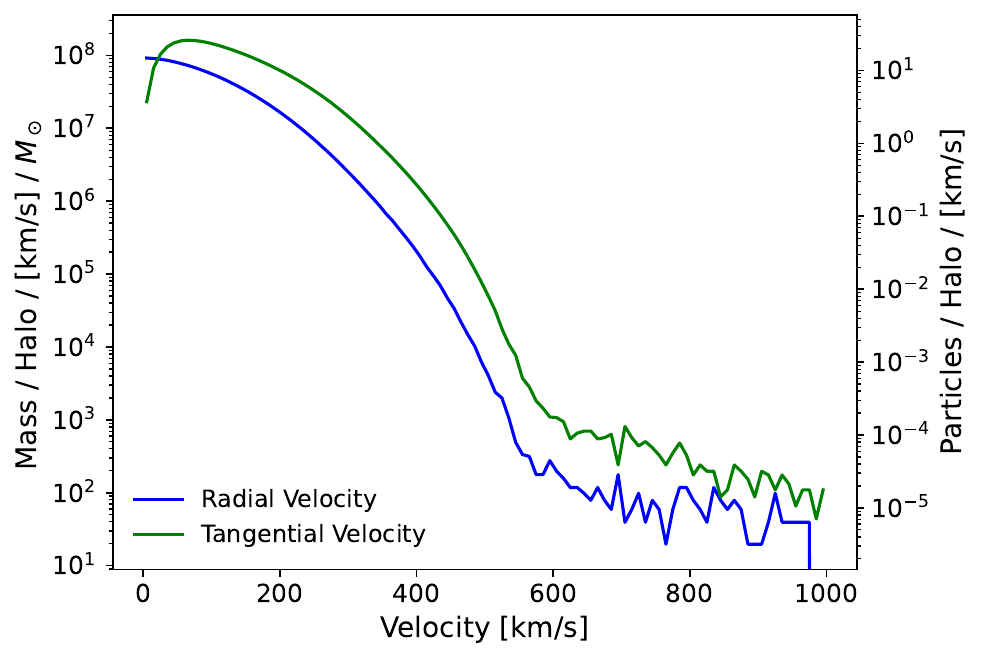}
    \caption{Distributions of radial and tangential velocities of DM particles in the Full catalog (i.e., central halos with $M_h=0.5-2\times 10^{12}\Msun$), within 5--11 kpc of the halo center.  The relatively higher tangential velocities imply an origin outside of the halo center (see text), reducing the chance of particle scattering being the origin of most high-velocity particles in the simulation.}
    \label{fig:radtang}
\end{figure}

The fact that the distribution of high-velocity particles is very non-Poissonian argues against the high-velocity particles being due to numerical noise in the simulation (e.g., unrealistically hard scattering of particles), as this is expected to be a Poisson process.  We can validate this intuition by analyzing the ratio between the tangential and radial velocity distributions, shown in Fig \ref{fig:radtang}. A higher frequency of tangential velocities is consistent with high velocity particles originating far from the center of the halo, as the ratio of tangential to radial motion would increase as particles move closer to the center of the halo due to the conservation of angular momentum. In contrast, particle collisions in the simulation would tend to produce high-velocity particles near the center of the halo due to the density--squared scaling for collision rates, giving preferentially lower angular momentum (radial) motion that is not seen in Fig.\ \ref{fig:radtang}.

High-velocity particles are also unlikely to arise from smooth accretion (defined as orbits largely determined by the MW host halo's potential instead of being influenced by another halo/subhalo), as by definition, the resulting particles would have velocities less than the escape velocity of the host halo.  Instead, high-velocity particles could plausibly come from: 1) massive nearby halos that have high velocity dispersions (leading to a flux of high-velocity particles through the MW halo), and/or 2) substructure that is orbiting or passing through the halo at high speeds.

\begin{figure}
	\includegraphics[width=\columnwidth,trim={0 0 0 0cm}, clip]{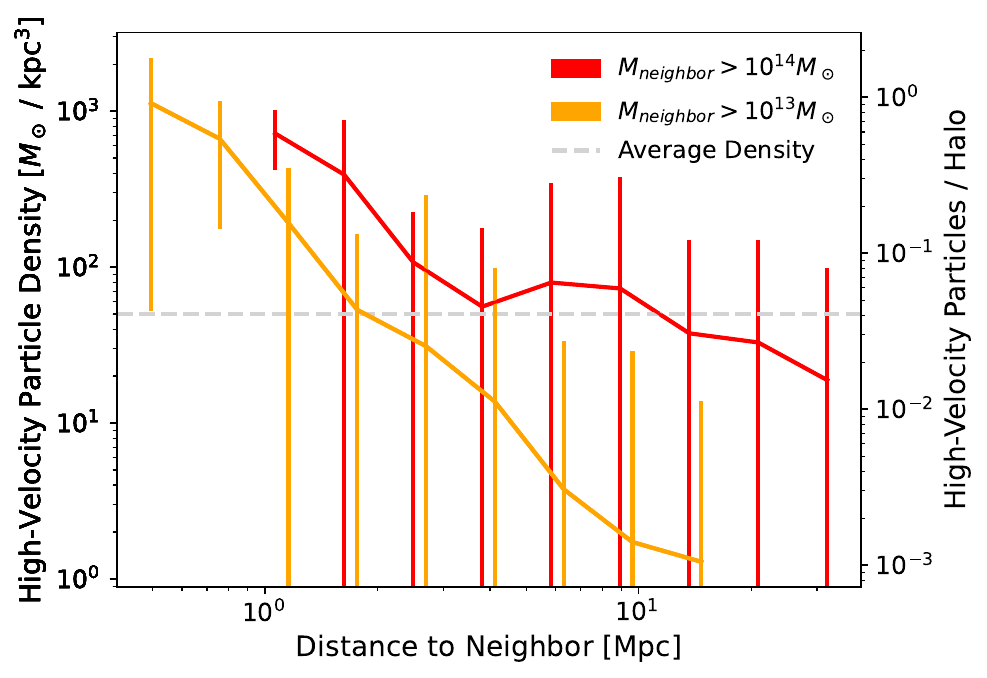}
    \caption{High velocity particle density near the solar neighborhood as a function of distance from MW-like halos to the nearest high-mass neighboring halos  ($>10^{14}\Msun$ and $>10^{13}\Msun$). This figure shows results for all central MW-like halos ($M_h=0.5-2\times 10^{12}\Msun$). Error bars show the  1-$\sigma$ dispersion across halos. Lastly, the average density of high-velocity particles across all MW-like halos is shown by the horizontal dashed line ($74.6\hspace{0.5mm} \Msun/$kpc$^{3}$).}
    \label{fig:hiM_hvpVSdist}
\end{figure}

\begin{figure}
	\includegraphics[width=\columnwidth,trim={0 0 0 0cm}, clip]{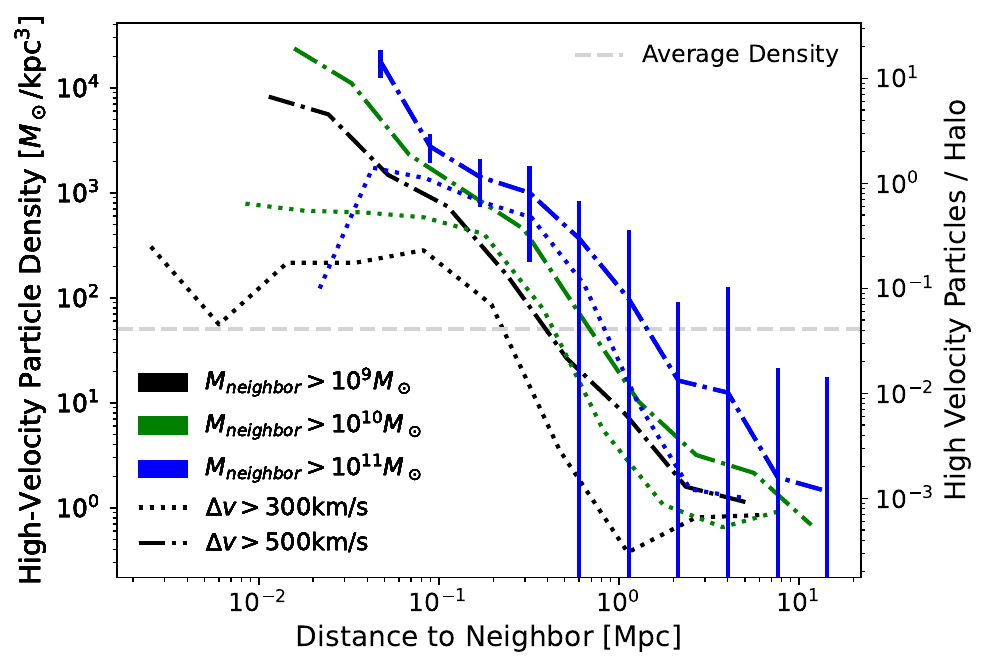}
    \caption{Central high velocity particle density within the solar neighborhood as a function of distance from MW-like halos to the nearest smaller neighboring halo above specified mass thresholds ($>10^{11}\Msun$, $>10^{10}\Msun$, and $>10^{9}\Msun$) and velocity thresholds ($\Delta v > 300$km/s and $\Delta v > 500$km/s).  Neighboring halos within the MW-like halos' virial radii ($\sim 300$ kpc) are satellite halos, whereas those beyond that radius are typically central halos.  Error bars show the  1--$\sigma$ dispersion across halos. Lastly, the average density of high-velocity particles across all MW-like  halos is shown by the horizontal dashed line ($74.6\hspace{0.5mm} \Msun/$kpc$^{3}$).} 
    \label{fig:loM_hiV_hvpVSdist}
\end{figure}

We investigate these two potential sources in 
Figures \ref{fig:hiM_hvpVSdist}--\ref{fig:loM_hiV_hvpVSdist}.
In all cases, halos have more high-velocity particles the closer their centers are to both massive neighboring halos and to high-velocity substructure.

Considering Fig.~\ref{fig:hiM_hvpVSdist} first, massive nearby halos can contribute substantially to high-velocity dark matter densities, up to a factor of $10\times$ higher than for average MW-like halos.  We find that massive halos increase the high-velocity particle density out to 3--4$\times$ their virial radii for both $>10^{13}$ and $>10^{14}\Msun$ halos.  This is because: 1) the turnaround radius of orbiting dark matter particles is $2-3$ virial radii, resulting in higher velocity dispersions beyond massive halos' virial radii; and 2) massive halos are highly clustered and so likely have nearby high-mass neighbors, which would also increase the velocity dispersion of dark matter particles in the same neighborhood.  However, the impact of a $>10^{14}\Msun$ cluster at the distance of Virgo is minimal.

Comparing Fig.\ \ref{fig:hiM_hvpVSdist} to Fig.\ \ref{fig:loM_hiV_hvpVSdist}, the impact of massive halos (up to $10\times$ higher high-velocity particle densities) is typically less than that of high velocity substructure, especially substructure moving close to the escape speed of the halo, which results in up to $100\times$ higher high-velocity particle densities. As expected, the faster the substructure, the more high velocity particles are present in the halo.  This is simply understood as the high-velocity substructure contributing directly to a localized excess in the particle velocity distribution.  This is evident from the mass-dependence of the high-velocity particle density excess, which is (as expected) greater for more massive substructure.

\begin{figure}
	\includegraphics[width=\columnwidth]{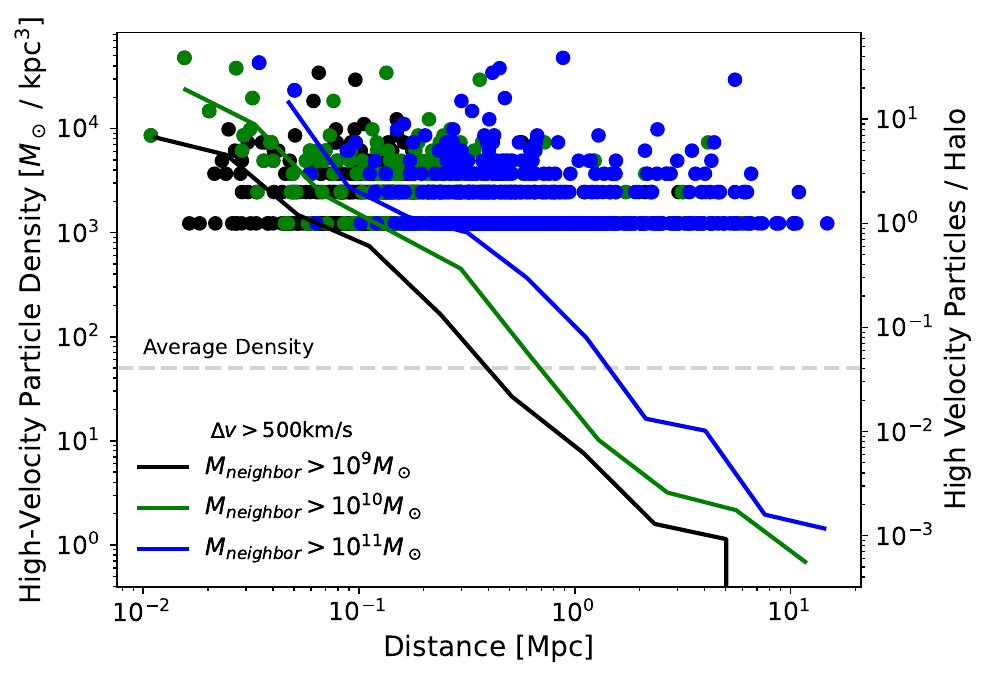}
    \caption{High velocity particle distribution within the solar neighborhood as a function of the distance from MW-like halos to the nearest high-velocity neighbor ($\Delta v > 500$ km/s) with $M_h>10^{9}\Msun$, $M_h>10^{10}\Msun$, $M_h>10^{11}\Msun$. The solid lines show the average across all halos with a neighbor at the given distance (as in Fig.\ \ref{fig:loM_hiV_hvpVSdist}), and the dots show high-velocity particle densities for individual halos. Lastly, the average density of high-velocity particles across all MW-like  halos is shown by the horizontal dashed line ($74.6\hspace{0.5mm} \Msun/$kpc$^{3}$)}
    \label{fig:hvpVSdist_Delta500}
\end{figure}

Fig.~\ref{fig:hvpVSdist_Delta500} shows the distribution of high-velocity particles in individual MW-like halos in the case that a neighboring smaller central or satellite halo has a high relative velocity ($\Delta v>500$ km/s). Not surprisingly, such cases lead to high incidences of high-velocity particles.  In particular, \textit{all} cases where a Milky Way like halo has a high-velocity particle density $>10^{4}\Msun$ kpc$^{-3}$ ($>200\times$ the average density) occur when there is substructure  with $\Delta v > 500$ km/s and a mass of $>10^{9}\Msun$.

\section{The Heliocentric Velocity Distribution in the Presence of the Large Magellanic Cloud}

\label{s:LMC}

  \begin{table}
    {\centering
    \caption{Expected boost to DM particle density for LMC-like substructure.}
    \label{tab:boost}
    \begin{tabular}{ccc}
    	\hline
    	Heliocentric $\Delta v$ & Density with LMC & Boost above average\\
    	\hline
$>600$ km/s & $3.0^{+0.2}_{-0.2}\times 10^4 \Msun\, \mathrm{kpc}^{-3} $ & +120\%\\
$>700$ km/s & $2.1^{+0.3}_{-0.3}\times 10^3 \Msun\, \mathrm{kpc}^{-3} $ & +270\%\\
$>800$ km/s & $1.7^{+0.7}_{-0.6}\times 10^2 \Msun\, \mathrm{kpc}^{-3} $ & +480\%\\
$>900$ km/s & $3.1^{+2.2}_{-1.6}\times 10^1 \Msun\, \mathrm{kpc}^{-3} $ & +110\%\\
     \hline
    	\end{tabular}
     }\\
     Selection: all Milky Way--like central halos ($0.5 < M_h / 10^{12}\Msun < 2$, distance to nearest $10^{14}\Msun$ halo at least 10 Mpc).  ``Density with LMC'' is calculated for those Milky-Way like central halos with an LMC-like subhalo ($\langle M_h \rangle = 10^{11.12}\Msun$) at $\langle D \rangle =50$ kpc with high relative heliocentric velocity ($470 < \Delta v_\odot < 570$ km/s).  Densities quoted ($\langle \rho_\mathrm{+LMC}\rangle$) are average dark matter densities at Earth-like distances (5--11 kpc) from the Milky Way analogs' halo centers.  ``Boost above average'' is calculated as $100\% \times (\langle \rho_\mathrm{+LMC}\rangle - \rho_\mathrm{all})/\rho_\mathrm{all}$, where $\rho_\mathrm{all}$ is the average over all Milky Way-like halos.  Errors correspond to  16--84$^\mathrm{th}$ percentile uncertainty ranges for the average density across all MW--LMC systems, computed by bootstrap resampling.
    \end{table}

While the previous section focused on the Galactocentric reference frame, it is straightforward to calculate the effect of the LMC on the DM distribution in a boosted heliocentric reference frame, which we do in this section.  Here, we adopt a heliocentric velocity of $\sim 250$ km/s with respect to the center of the Milky Way, following \cite{Kallivayalil13}.  To estimate the heliocentric velocity distribution for typical halos, we apply boosts of $250$ km/s along 6 different directions in the \textit{VSMDPL} simulation ($\pm$ X, $\pm$ Y, and $\pm$ Z), and average the resulting particle velocity distributions for MW-like halos in the Full catalog over all boosted frames.
    
We select MW-like halos with LMC-like objects based on observed constraints for the LMC and its dynamics.  From the Full catalog, we select objects with a subhalo with $M_h>10^{10.5}\Msun$ (based on an inferred LMC mass of $10^{11.11\pm 0.3}\Msun$; \citealt{Vasiliev21}) at distances of $33-73$ kpc (based on an inferred distance to the LMC of 50 kpc; \citealt{LMC_Dist}).  We then measure the particle velocity distributions at 5--11~kpc from the host halo center for those MW--LMC analogs that have relative velocities of $470$ km/s $< \Delta v < $ 570~km/s in one or more of the boosted frames above (based on an observed relative velocity of 520 km/s; \citealt{Kallivayalil13}).  We average the particle velocity distribution over all 487 frame--halo pairs that pass the selection cuts above, and compare to the average boosted (heliocentric) particle velocity distributions for all MW-like halos from the Full catalog.  Of note, the average halo mass of the LMC-like objects that pass all selection cuts above is $10^{11.12}\Msun$, and the average halo mass of their MW-like hosts is $10^{12.13}\Msun$.

We find that the presence of an LMC-like object strongly correlates with enhanced DM particle density at Earth-like distances, as shown in Table \ref{tab:boost}.  LMC-correlated boosts increase from 120\% higher particle densities at $\Delta v>600$ km/s to 480\% higher particle densities at $\Delta v>800$ km/s.  The LMC-correlated contribution, while significant, declines again above 900 km/s, approximately the sum of our upper relative velocity threshold (570 km/s) and the median escape velocity of the LMC-like subhalos ($\sim 300$ km/s).

\section{Discussion}
\label{s:discussion}
Most halos do not have significant mass in high velocity dark matter particles. Indeed, for most Milky Way-like halos, we find a halocentric velocity distribution of DM particles truncated at around 600 km/s. This is consistent with the findings of \cite{Santos23} where it was determined the local maximum DM speed in the Solar Neighborhood is approximately 597 km/s, as well as earlier papers on the MW velocity distribution (e.g., \citealt{Mao13}).
Nonetheless, current methods for detecting DM particles are preferentially sensitive to the highest energy particles. By analyzing a large-volume cosmological dark matter simulation with many Milky Way like halos, we were able to better characterize the origin of these particles when they occur. Our results show that the presence of high velocity DM particles is enhanced in cases where the host halo has a massive neighboring halo but even more so when it has nearby fast moving substructure.

The most likely candidate previously advanced for increasing the expected high velocity DM particle distribution is the Large Magellanic Cloud (LMC). This was suggested by zoom-in and isolated simulations conducted by \cite{Orlik23} and \cite{Besla19}. Both papers found that the presence of the LMC results in a significant influence on the local DM distribution, with \cite{Besla19} concluding that the high speed tail of the MW dark matter distribution from 700--900 km/s is ``overwhelmingly of LMC origin.''  We confirm this assessment, finding in our analysis that nearly 75\% of the particles traveling at these speeds in the heliocentric reference frame are correlated with the presence of an LMC-like object (Table \ref{tab:boost}).

  Last, we also found that the presence of M31 near the MW does not significantly influence the expected number of high velocity particles. The lack of influence of M31 on the velocity distribution is similar to the finding in \cite{Behroozi13}, who concluded that the presence of M31 was not likely to impact direct detection experiments because it does not increase the overall number of particles within the halo radius beyond the escape velocity (albeit using a much lower-resolution simulation).


\section{Conclusions} 
\label{s:conclusions}
In this paper, we examined the central velocity distribution for dark matter in Milky Way--like halos.  Our main results are:
\begin{itemize}
    \item For the average Milky Way--mass halo, a fraction $\sim 1.3\times 10^{-5}$ of the DM mass near the Earth's position is high-velocity ($\Delta v>600$ km/s; Section \ref{s:results}). 
    \item The presence of a nearby massive $M_h>10^{13}\Msun$ halo can increase the expected number of high-velocity particles by an order of magnitude (Section \ref{s:results}, Fig.\ \ref{fig:hiM_hvpVSdist}).
    \item The presence of a nearby high-velocity substructure can increase the expected number of high-velocity particles by two orders of magnitude (Section \ref{s:results}, Fig.\ \ref{fig:loM_hiV_hvpVSdist}). 
    \item The Large Magellanic Cloud is expected to substantially boost high-velocity particle density in the Milky Way, especially for heliocentric velocities $>$700 km/s (Section \ref{s:LMC}, Table \ref{tab:boost}).
    \item The presence of an M31-like neighbor does not impact the number of high-velocity particles beyond 600 km/s (Section \ref{s:results}, Fig.\ \ref{fig:hiM_hvpVSdist}).

\end{itemize}

\section*{Acknowledgements}
We thank Gurtina Besla, Haley Bowden, Katie Chamberlain, Hayden Foote, Elaheh Hayati, Annika Peter, Himansh Rathore, and  Haowen Zhang for their insight and contributions.  PB was funded by a Packard Fellowship, Grant \#2019-69646.  The authors gratefully acknowledge the Gauss Centre for Supercomputing e.V. (www.gauss-centre.eu) and the Partnership for Advanced Supercomputing in Europe (PRACE, www.prace-ri.eu) for funding the MultiDark simulation project by providing computing time on the GCS Supercomputer SuperMUC at Leibniz Supercomputing Centre (LRZ, www.lrz.de). 




\bibliographystyle{mnras}
\bibliography{references} 







\label{lastpage}
\end{document}